\documentstyle[prl,aps,epsf,preprint]{revtex}
\begin{document}
\tighten

\title{LORENTZ SYMMETRY AND \\
THE INTERNAL STRUCTURE OF THE NUCLEON}

\author{Xiangdong Ji }

\address{Department of Physics \\
University of Maryland \\
College Park, Maryland 20742}

\date{U. Md PP\# 98-042,~~ DOE/ER/40762-133,~~ September 1997}
\maketitle

\begin{abstract}
To investigate the internal structure of the 
nucleon, it is useful to introduce quantities
that do not transform properly under 
Lorentz symmetry, such as the four-momentum of the quarks in 
the nucleon, the amount of the nucleon
spin contributed by quark spin, etc.  In this paper,
we discuss to what extent these
quantities do provide Lorentz-invariant descriptions
of the nucleon structure. 
\end{abstract}

\pacs{xxxxxx}

\narrowtext
In field theory, one often encounters 
various {\it densities} consisting of elementary fields,
space-time coordinates and their derivatives:
baryon current, momentum density, angular momentum
density, etc. These densities are Lorentz
covariant, i.e., under Lorentz tranformations,
they transform properly as four-vectors or
four-tensors. In many cases, we are interested also
in the {\it charges} defined from these densities. 
Considering a generic density $j^{\mu\alpha...}$, one can 
define a charge according to, 
\begin{equation}
       Q^{\alpha...} = \int d^3x j^{0\alpha...}\ . 
\end{equation}
Generally speaking, $Q^{\alpha...}$ no longer
transforms properly under Lorentz transformations.
The condition for $Q^{\alpha...}$ to be Lorentz
covariant is well known \cite{weinberg}: The density $j^{\mu\alpha...}$
must be conserved relative to the index $\mu$, 
\begin{equation}
      \partial_\mu j^{\mu\alpha...} = 0 \ . 
\end{equation}
Indeed in most of the applications, one  
considers charges from conserved densities. 

Nevertheless, it is useful to consider
charges defined from non-conserved densities. 
For instance, in quantum chromodynamics (QCD), 
the energy-momentum density consists of 
the sum of quark and gluon contributions,
\begin{equation}
     T^{\mu\nu} = T^{\mu\nu}_q + T^{\mu\nu}_g \ . 
\end{equation}
The total density is conserved due to translational
invariance,
$\partial_\mu T^{\mu\nu} = 0$. Therefore,
the total momentum operator $P^{\mu}$,
\begin{equation}
      P^{\mu} = \int d^3 x T^{0\mu} \ , 
\end{equation}
tranforms like a four-vector under Lorentz symmetry.
Meanwhile, one can also introduce the notion of 
the four-momenta carried separately 
by quarks and gluons,
\begin{equation}
       P^{\mu}_{q,g}(\mu) = \int d^3x T^{0\mu}_{q, g} \ , 
\end{equation}
where the non-conservation of $T^{0\mu}_{q,g}$
calls for a renormalization scale $\mu$. 
It is quite obvious that $P^{\mu}_{q,g}$
do not transform like four-vectors, and therefore
significance of such quantities appears doubtful.
However, the exact transformation property of 
the expectation values of $P^\mu_{q,g}$ is simple to derive.

The forward matrix element of the total energy-momentum
density in a nucleon state is,
\begin{equation}
      \langle p|T^{\mu\nu}|p\rangle
    = 2p^\mu p^\nu \ . 
\label{emt}
\end{equation}
Here the covariant normalization 
of the nucleon state is used. From the above, one
can easily obtain the usual matrix element of the 
momentum operator. The matrix elements of 
quark and gluon parts of the density involve
two Lorentz structures,
\begin{equation}
      \langle p |T^{\mu\nu}_{q,g}| p\rangle
      = 2A_{q,g}(\mu)p^{\mu}p^{\nu} + 2B_{q,g}(\mu)g^{\mu\nu}
\end{equation}
where $A_{q,g}$ and $B_{q,g}$ are scalar constants.
Comparing with Eq. (\ref{emt}), one has
the following constraints,
\begin{eqnarray} 
      A_q(\mu) + A_g(\mu) &=& 1 \ , \nonumber \\
      B_q(\mu) + B_g(\mu) &=& 0 \ . 
\end{eqnarray}
Moreover, the quark and gluon contributions
to the nucleon four-momentum are, 
\begin{equation}
       \langle p | P^\mu_{q,g}|p \rangle
      = A_{q,g}(\mu)p^\mu + B_{q,g}(\mu)g^{\mu 0}/(2p^0) \ . 
\label{mom}
\end{equation}
The above equation defines transformation
properties of the expectation values of $P^\mu_{q,g}$. 
The presence of the second term denies them a proper
Lorentz transformation. 

On the other hand, if one is interested in the {\it three-momentum} of the
nucleon only, the second term in Eq. (\ref{mom}) drops
out and three components of the matrix
elements transform just like those 
of a four-vector. Because $A_{q,g}(\mu)$ are Lorentz
scalars, one concludes that the fractions of the
nucleon three-momentum carried by quarks
and gluons are invariant under Lorentz tranformations!
Such a statement, although drawn for non-Lorentz-covariant
quantities, does carry important physical significance.
Phenomenologically, $A_{q,g}(\mu)$ have been extracted 
from the parton distributions which have simple 
interpretations only in the infinite momentum frame. 
According to the above discussion, the fractions of 
the nucleon mometum carried by 
quarks and gluons are also the same in ordinary frames. 
In particular, if a nucleon 
has a momentum of 1 GeV/c, then according to the
recent analysis \cite{cteq}, roughly 420 MeV/c  
is carried by gluons in the form of the Poynting vector
$\int d^3x  \vec{E}\times \vec{B}$ in the $\rm\overline{MS}$ scheme
and at $\mu$ = 1.6 GeV. 

A more intriguing example concerns the spin 
structure of the nucleon, which depends on 
the QCD angular momentum density. The total 
density $M^{\mu\alpha\beta}$ is a mixed 
Lorentz tensor, expressible in terms of the 
energy-momentum density $T^{\mu\nu}$ in the 
Belinfante form \cite{belinfante,jaffe},
\begin{equation}
    M^{\mu\alpha\beta} 
    = T^{\mu\beta} x^\alpha - T^{\mu\alpha} x^\beta \ . 
\label{amd}
\end{equation}
The relevant charges $J^{\alpha\beta} = 
\int d^3x M^{0\alpha\beta}$ are the usual Lorentz
generators (including the angular momentum operator
$\vec{J}$), which tranform as the Lorentz
tensor $(1,0)+(0,1)$.

According to
Eq. (\ref{amd}), the angular momentum 
density has both quark and gluon contributions,
$M^{\mu\alpha\beta} = M^{\mu\alpha\beta}_q +
M^{\mu\alpha\beta}_g$, where,
\begin{equation}
         M^{\mu\alpha\beta}_{q,g} = T^{\mu\beta}_{q,g}x^\alpha
      - T^{\mu\alpha}_{q,g}x^\beta \ . 
\label{aa}
\end{equation}
Furthermore, the quark part contains the 
spin and orbital contributions,
        $M^{\mu\alpha\beta}_q = M^{\mu\alpha\beta}_{qL}
            + M^{\mu\alpha\beta}_{qS}$,
where
\begin{eqnarray}
  M^{\mu\alpha\beta}_{qS} &= &{i\over 4}\bar \psi
    \gamma^\mu[\gamma^\alpha,\gamma^\beta] \psi \ ,  \nonumber \\
  M^{\mu\alpha\beta}_{qL} & = & \bar \psi \gamma^\mu (x^\alpha iD^\beta
     - x^\beta iD^\alpha)\psi \ .
\label{bb}
\end{eqnarray}
Accordingly, the QCD angular momentum operator 
can be written as a sum of three gauge-invariant 
contributions,  
\begin{eqnarray}
     \vec{J} &= &\int d^3x \psi^\dagger {\vec{\Sigma}\over 2}
           \psi + \int d^3x \psi^\dagger (\vec{x}
           \times (-i\vec{D})\psi
         + \int d^3x \vec{E}\times \vec{B} \nonumber \\
      &\equiv & \vec{S_q} + \vec{L_q} + \vec{J_g} \ . 
\label{ang}
\end{eqnarray}                      
An individual term in the above expression
does not transform as a part of $(1,0)+(0,1)$,  
and does not satisfy the angular 
momentum algebra by itself. 
Nonetheless, as we will argue, the decomposition 
is useful in studying the spin 
structure of the nucleon.

The angular momentum operator in Eq. (\ref{ang})  
not only generates the spin of a composite particle,
but also describes the orbital motion of its center 
of mass. To separate the two, Pauli and Lubanski
introduced a spin vector $W^\mu$ \cite{lubanski}, 
\begin{equation}
       W^{\mu} = -\epsilon^{\mu\alpha\beta\gamma}
              J_{\alpha\beta} P_\gamma/(2\sqrt{P^2}) \ ,  
\end{equation}
which reduces to the angular momentum operator
only in the rest frame of the particle.
The spin quantum number $s$ characterizes  
the eigenvalue $s(s+1)$ of the scalar 
Casimir $W^\mu W_\mu$ in a representation 
of the Poincare algebra. The separation
of internal and external motions in a general 
Lorentz frame comes with a price: $W^\mu$ 
contains not only the 
angular momentum operators but also the boost operators. 
Consequently, the scalar character (or invariant notion) 
of the spin, 
\begin{equation}
      \langle p|W^\mu W_\mu|p\rangle /\langle p|p \rangle = 
      {1\over 2}({1\over 2}+1) \ .  
\end{equation} 
is maintained in an elaborated way in different
reference frames. The above
equation does not offer a convenient starting point
to investigate the spin structure of the nucleon. 

To specify the spin states of a nucleon, 
a polarization vector $s^\mu$ is usually introduced
\cite{bd}. In the rest frame of the nucleon,
$s^{\mu}$ specifies the direction of the spin 
quantization axis; the convention in the literature
is such that $s^\mu$ represents a state with 
positive spin projection 1/2. Thus, in any Lorentz 
frame, 
\begin{equation}
     \langle ps|W^\mu s_\mu|ps \rangle/
   \langle ps|ps \rangle    = {1\over 2} \ . 
\end{equation}
The above equation is linear in angular momentum operators
and is suitable for studying the spin structure if  
the boost operators are not present. This is the case
for a special choice of $s^\mu = (|\vec{p}|/M,p^0\hat p/M)$, 
where $\hat p = \vec{p}/|\vec{p}|$. $W^\mu s_\mu$ then 
is just the well-known 
helicity operator $h = \vec{J}\cdot \hat p$ and the above 
equation reduces to,  
\begin{equation}
     \langle p + | J\cdot \hat p| p +\rangle
   /\langle p +|p+\rangle 
     = {1\over 2} \ , 
\label{hel}
\end{equation}
where + denotes the positive helicity. Because of 
the special choice, Eq. (\ref{hel})
is invariant only under a special class
of Lorentz transformations, i.e, the boosts along
the momentum without reversing its direction 
and rotations around the momentum axis. 
While the conclusion is quite 
obvious, our analysis shows that the study of
the spin structure can at best
be done in a restricted class of Lorentz 
frames. 

In the remainder of the paper, we are going to show
that when the angular momentum operator
is split into a sum as in Eq. (\ref{ang}),
the individual contributions to the spin 
of the nucleon are invariant under the special 
Lorentz transformations that preserve the 
helicity.

Without loss of generality, let us assume that 
the nucleon momentum is in the positive $z$ direction. 
Then the helicity operator is just $J^z = J^{12} 
= \int d^3 x M^{012}$. 
Consider the matrix elements of 
$\int d^4x M^{\mu\alpha\beta}$ in the nucleon state,
\begin{eqnarray}
    \left\langle p \left| \int d^4x M^{\mu\alpha\beta} \right|p
     \right\rangle 
    &=& \bar U(p) \Big[{i\over 2}A\left(p^\mu[\gamma^\alpha,\gamma^\beta]
    + \gamma^\mu[\gamma^\beta p^\alpha - \gamma^\alpha p^\beta]\right)
    \nonumber \\
    && + B \left( g^{\mu\alpha}\gamma^\beta 
  - g^{\mu\beta}\gamma^\alpha
    \right)\Big]U(p)(2\pi)^4\delta^4(0) + ...
\label{matrix}
\end{eqnarray}
where $...$ denotes terms with derivatives on the 
$\delta$ function which are related to the orbital 
motion. Taking $\mu=0$, $\alpha=1$, and $\beta=2$, we have,
\begin{equation}
      \langle p +|J^z| p + \rangle /\langle
        p+|p+\rangle = A \ .  
\end{equation}
Thus $A$ must be 1/2. Now suppose $M^{\mu\alpha\beta} = \sum_i 
M^{\mu\alpha\beta}_i$. The Lorentz symmetry yields,
\begin{eqnarray}
    \left\langle p \left| \int d^4x M^{\mu\alpha\beta}_i 
    \right|p \right\rangle
    &=& \bar U(p) \Big[{i\over 2}A_i(\mu) p^\mu[\gamma^\alpha,\gamma^\beta]
    + B_i(\mu)\gamma^\mu\left(
  \gamma^\beta p^\alpha - \gamma^\alpha p^\beta\right)
    \nonumber \\
    && + C_i(\mu) \left(g^{\mu\alpha}\gamma^\beta 
  - g^{\mu\beta}\gamma^\alpha\right)\Big] U(p)(2\pi)^4\delta^4(0) + ...
\label{hi}
\end{eqnarray}
where $A_i(\mu)$, $B_i(\mu)$ and $C_i(\mu)$ are scalar constants. 
Then the consistency between Eqs. (\ref{matrix}) and (\ref{hi}) yields, 
\begin{equation}
      \sum_i A_i(\mu) = A = {1\over 2}\ . 
\label{sum}
\end{equation}

Equation (\ref{hi}) allows us to calculate the 
contributions to the nucleon spin from different
terms in $\vec{J} = \sum_i \vec{J}_i$, 
\begin{equation}
     \langle p +| J^z_i | p + \rangle/\langle p+|p+\rangle 
     = A_i(\mu) \ , 
\end{equation}
where $J^z_i = \int d^3x M^{012}_i$. 
The result is {\it independent} of the 
magnitude of the nucleon momentum. Futhermore, 
Eq. (\ref{sum}) yields a nucleon spin
sum rule invariant under the special Lorentz 
transformations.
Denoting the scalar matrix elements $A_i$ 
of the angular momentum densities 
in Eqs. (\ref{aa},\ref{bb})
as $A_{qS} \equiv {1\over 2}\Delta \Sigma$, 
$A_{qL} \equiv L_q$ and $A_g \equiv J_g$, Eq. (\ref{sum})
becomes
\cite{ji}, 
\begin{equation}
      {1\over 2}\Delta \Sigma(\mu) + L_q(\mu) + J_g(\mu) = {1\over 2} \ . 
\end{equation}
As the nucleon momentum goes to infinity, 
the result can be interpreted
according to light-front quantization, in which the 
Lorentz generators are defined according to
$\int d^3x M^{+\alpha\beta}$. As such, the quark 
spin and orbital and gluon angular momentum 
contributions to the nucleon are the same 
in both ordinary and light-front coordinates.

\acknowledgments
The author wish to thank R. L. Jaffe for questioning
and discussing the invariance
of the spin structure of the nucleon, and T. Cohen, 
P. Hoodbhoy, and W. O. Greenberg for useful conversations.
Support of the U.S. Department of Energy under grant
number DE-93ER-40762 is gratefully acknowledged.


\begin{references}
\bigskip
\frenchspacing

\bibitem{weinberg}
S. Weinberg, {\it Gravitation and Cosmology} (Wiley, New York, 1972.)

\bibitem{cteq}
H. L. Lai et al., hep-ph/9606399, 1996. 

\bibitem{belinfante}
F. Belinfante, Physica {\bf 7}, 449 (1940). 

\bibitem{jaffe}
R. L. Jaffe and A. Manohar, Nucl. Phys. B337 (1990) 509. 

\bibitem{lubanski}
J. K. Lubanski, Physica {\bf 9}, 310 (1942). 

\bibitem{bd}
J. D. Bjorken and S. D. Drell, {\it Relativistic Quantum Mechanics}
(McGraw-Hill, New York, 1964.) 

\bibitem{ji}
X. Ji, Phys. Rev. Lett. {\bf 78}, 610 (1997).

\nonfrenchspacing
\end{references}
\end{document}